\documentclass[a4paper,12pt,reqno]{amsart}
\usepackage{latexsym}
\usepackage{amsmath}
\usepackage{amssymb}
\usepackage{mathrsfs}
\usepackage[all]{xy}
\usepackage{appendix}
\usepackage[pdftex]{graphicx}
\usepackage[dvips,letterpaper,margin=0.9in]{geometry}
\usepackage{braket}
\usepackage{setspace}
\usepackage{booktabs}
\usepackage{enumerate}
\usepackage{pdflscape}
\usepackage{url}
\usepackage[bottom]{footmisc}
\usepackage{tikz}

\usepackage{hyperref}
\hypersetup{
    colorlinks,
    citecolor=red,
    filecolor=black,
    linkcolor=blue,
    urlcolor=black
}

\begin{document}

\title{Conformal gravity as a deformed topological field theory}
\author{James A. Reid}
\address[James A. Reid]{Darkocean, Office 9, Building 2, Financial Square, Doha, Qatar}
\email{j.reid.06@aberdeen.ac.uk}
\date{\today}

\newcommand{\onetoone}{\overset{1:1}{\iff}}
\newcommand{\ud}{\mathrm{d}}
\newcommand{\wt}{\widetilde}
\newcommand{\wh}{\widehat}
\newcommand{\tr}{\text{tr~}}
\newcommand{\p}{\phantom}
\newcommand{\la}{\langle}
\newcommand{\ra}{\rangle}
\newcommand{\wb}{\overline}
\newcommand{\contract}{\lrcorner}
\newcommand{\dbar}{\overline{\partial}}
\newcommand{\head}[1]{\textnormal{\textbf{#1}}}
\newcommand{\poincare}{Poincar\'{e}}
\newcommand{\pe}{Poincar\'{e}-Einstein }
\newcommand{\poinein}{Poincar\'{e}-Einstein}
\newcommand{\zeroinfty}{[0, \infty)}
\newcommand{\tickbox}{\makebox[0pt][l]{$\square$}\raisebox{.15ex}{\hspace{0.1em}$\checkmark$}}
\newcommand{\longhookrightarrow}{\lhook\joinrel\longrightarrow}

\newtheorem{definition}{Definition}
\newtheorem{lemma}{Lemma}
\newtheorem{theorem}{Theorem}
\newtheorem{corollary}{Corollary}
\newtheorem{calculation}{Calculation}
\newtheorem{proposition}{Proposition}
\newtheorem{fact}{Fact}
\newtheorem{mainresult}{Main Result}
\newtheorem{identity}{Identity}
\newtheorem{speculation}{Speculation}

\newcommand{\cone}{\mbox{\textcircled{\small 1}}}
\newcommand{\ctwo}{\mbox{\textcircled{\small 2}}}
\newcommand{\cthree}{\mbox{\textcircled{\small 3}}}
\newcommand{\cfour}{\mbox{\textcircled{\small 4}}}
\newcommand{\cfive}{\mbox{\textcircled{\small 5}}}

\newcommand{\circex}{\mbox{\textcircled{\small \boldsymbol{!}}}}
\newcommand{\circa}{\mbox{\textcircled{\small a}}}
\newcommand{\circb}{\mbox{\textcircled{\small b}}}
\newcommand{\circc}{\mbox{\textcircled{\small c}}}
\newcommand{\circd}{\mbox{\textcircled{\small d}}}

\begin{abstract}
In the MacDowell-Mansouri formulation of general relativity, the spin connection and coframe variables are incorporated into a single Lie algebra-valued connection called the MacDowell-Mansouri connection, $\omega$.  From the curvature form $F$ of $\omega$ and an auxiliary field, $B$, one may formulate general relativity as a deformed topological field theory by constructing an action functional whose variation yields a set of field equations that are equivalent to the Einstein equations on shell.  In this article, we show that when the fundamental length scale of the MacDowell-Mansouri connection is regarded as a dynamical variable - a cosmological scalar field - the field equations obtained from the variation of the resulting action are equivalent to the conformal Einstein equations on shell.  Through the lens of Cartan geometry, we then discuss a notable geometrical difference between general relativity and its conformally transformed counterpart.  Specifically, for the latter, we show that points in spacetime are infinitesimally approximated by homogeneous spaces (restricted to a point) whose radii are parameterised by the value of the cosmological scalar field.\\ \\
PACS numbers: 04.50.Kd, 04.20.-q, 02.40.-k.\\
\flushright \textit{On the occasion of Sir Roger Penrose's 94\textsuperscript{th} birthday.}
\end{abstract}
\maketitle
%
%

\section{Motivation}
In fundamental physics, no mathematical concept is arguably more important, or more far-reaching in its consequences, than that of the action functional.  By varying an action functional, one obtains in a straightforward manner the field equations of the theory under consideration.  It is little wonder then that the cover of the storied \emph{Encyclopedia of Mathematical Physics} is emblazoned with the solitary, bold expression
\begin{equation*}
\delta \int \mathcal{L} = 0,
\end{equation*}
articulating with brevity its ubiquity in physical theories of nature.  The importance of action functionals in general relativity is perhaps attested to by the multitude that have been constructed.  Other than the Einstein-Hilbert, MacDowell-Mansouri and Palatini actions that we will discuss in this article, other examples include those constructed by Arnowitt-Deser-Misner \cite{adm}, Baierlein-Sharp-Wheeler \cite{bsw}, Capovilla-Dell-Jacobson-Mason \cite{capovilla}, Goldberg \cite{goldberg}, Pleb\'{a}nski \cite{plebanski} and numerous others (\emph{c.f.} \cite{bombelli} for a large list).  Similarly, there are almost as many action functionals for scalar-tensor theory: gravitational theories that incorporate in a fundamental way, a cosmological scalar field.  In this article we will construct a deformed topological field theory action for conformally transformed general relativity, and show that the cosmological scalar field therein has a direct geometric interpretation in the context of Cartan geometry. Conformally transformed general relativity is commonly known as conformal gravity, or scale-invariant gravity \cite{kelleher, moon}, and the familiar action functional from which the field equations are derived is called the conformal Einstein-Hilbert action \cite{diraclongrange}:
\begin{equation}\label{eq:cehaction}
S_{\text{CEH}} :=  \int_{M} \ud^{4} x~ \sqrt{-g}  \Big( \phi^{2} \mathsf{scal}(\{ ~ \})  + 6 g^{ab} \partial_{a} \phi \partial_{b} \phi - 2 \Lambda \phi^{4} \Big), 
\end{equation}
where $M$ is a smooth 4-manifold with boundary, $\phi$ the cosmological scalar field and $\{ ~ \}$ denotes the Christoffel symbol of the Levi-\v{C}ivit\`{a} connection. This action is so-called because it is the image of the conformal transformation $g_{ab} \rightarrow e^{2 \Upsilon} g_{ab}$ of the Einstein-Hilbert action with a cosmological constant:
\begin{equation}
S_{\text{EH}} = \int_{M} \ud^{4} x~ \sqrt{-g} \Big( \mathsf{scal}(\{ ~ \}) - 2 \Lambda \Big),
\end{equation}
where $\phi = e^{ \Upsilon}$ for $\Upsilon$ a smooth function. Varying the conformal Einstein-Hilbert action, one obtains the (vacuum) conformal Einstein equations
\begin{equation}\label{eq:cefestogether}
\bigg\{ \begin{array}{l}
0 = \phi^{2} \mathsf{G}_{ab}(\{~ \}) + 4 \partial_{a} \phi \partial_{b} \phi - g_{ab} \partial_{c} \phi \partial^{c} \phi - 2 \phi \left( \partial_{a} \partial_{b}  - g_{ab} \square \right) \phi + g_{ab} \Lambda \phi^{4},\\
0 = 2 \phi \mathsf{scal}(\{~ \}) - 12 \square \phi - 8 \Lambda \phi^{3},
\end{array}
\end{equation}
where
\begin{equation}
\mathsf{G}_{ab}(\{~ \}) = \mathsf{Ric}_{ab}(\{ ~ \}) - \frac{1}{2} \mathsf{scal}(\{ ~ \}) g_{ab}
\end{equation}
is the Einstein tensor (of the Levi-\v{C}ivit\`{a} connection).  Conformal gravity is so-called because both the conformal Einstein-Hilbert action \eqref{eq:cehaction} and the conformal Einstein equations \eqref{eq:cefestogether} are invariant with respect to the joint conformal transformations\footnote{In addition to these facts, one may additionally show that the vacuum conformal Einstein equations are the image of the conformal transformation of the vacuum Einstein equations.} $g_{ab} \rightarrow e^{2 \Upsilon} g_{ab}$ and $\phi \rightarrow e^{- \Upsilon} \phi$. This fact was one of Dirac's motivations for basing an alternative cosmology on conformal gravity: he believed that ``basic [physical] laws should be invariant under the widest possible group of transformations'' \cite{diraclongrange}.  Another motivation for conformal gravity stems from canonical general relativity.  It has been argued that the dynamical degrees of freedom of the gravitational field ought to be identified with conformal 3-geometries, which are to say, equivalence classes of conformal metrics upto diffeomorphism \cite{york}. Consequently, the configuration space of general relativity should not be identified with superspace\footnote{The superspace \cite{fischersuperspace} of a closed 3-manifold $M$ is the orbit space $\mathscr{S}(M) := \text{Riem}(M)/\text{Diff}(M)$ of the diffeomorphism group acting by coordinate transformation on the space of Riemannian metrics on $M$, $\text{Riem}(M)$.}, but rather with conformal superspace: the space of Riemannian metrics modulo diffeomorphisms and conformal transformations (\emph{c.f.} \cite{fischer} for a precise definition).  The conformal Einstein-Hilbert action has been shown to decompose to a canonical gravity action describing a theory whose configuration space is precisely conformal superspace \cite{kelleher}.  The conformal Einstein-Hilbert action has also been adopted by Mannheim as a conformally invariant matter action for his conformal cosmology \cite{mannheim}.\\

Our construction of an action for conformal gravity however, is motivated by providing a geometrical description for the scalar field $\phi$.  Indeed, one of the proposed criticisms of scalar-tensor theories of gravity in general is that the cosmological scalar field does not have a geometrical origin \cite{faraoni}.  In the early 1950's, Lyra addressed this pathology by formulating scalar-tensor gravity on (the now eponymous) Lyra manifolds \cite{lyra}.  A Lyra manifold is a triple $(M,g,\psi)$ consisting of an $n$-dimensional manifold $M$ with semi-Riemannian metric $g$ and a scalar field $\psi$ called the gauge function.  Lyra's connection coefficients take the form \cite{faraoni}
\begin{equation}
\Gamma^{c}_{\p{c}ab} = \frac{1}{\psi}  \left\{ \begin{array}{c} c \\ a ~~~ b  \end{array} \right\} + \frac{s+1}{\psi^{2}} g^{cd} \left( g_{bd} \partial_{a} \psi - g_{ab} \partial_{d} \psi \right),
\end{equation}
where $s$ is a constant related to the torsion of Lyra's connection.  The action for the 4-dimensional Lorentzian theory,
\begin{equation}
S_{\text{Lyra}} = \int_{M} \ud^{4} x~ \sqrt{-g}~ \psi^{4} \mathsf{K},
\end{equation}
with $\mathsf{K}$ the Lyra scalar curvature \cite{soleng}
\begin{equation}
\mathsf{K} = \frac{\mathsf{scal}}{\psi^{2}} + \frac{2(s+1)}{\psi^{3}} (1-n) \square \psi + \frac{1}{\psi^{4}} \Big[ (s+1)^{2} \left( 3n - n^{2} - 2 \right) - (s+1)(2-2n) \Big] \partial_{a} \psi \partial^{a} \psi,
\end{equation}
has been shown to be equivalent to the scalar-tensor theory given by the action
\begin{equation}\label{eq:lyrabdaction}
S = \int_{M} \ud^{4} x~ \sqrt{-g}~ \left( \psi^{2} \mathsf{scal} - 4 \omega g^{ab} \partial_{a} \psi \partial_{b} \psi \right)
\end{equation}
in $n=4$ dimensions, where the coupling constant $\omega$ is related to the torsional constant $s$ according to $\omega = 3(s^{2}-1)/2$. Other formulations of scalar-tensor gravity in which the scalar field has a geometrical interpretation include Kaluza-Klein theory and bosonic string theory.  In the former, the scalar field is the determinant of the metric on the $d$ extra spatial dimensions and in the latter, the scalar field is the dimensionless string dilaton in the low energy action in the string frame \cite{faraoni}.

\section{MacDowell-Mansouri Gravity and Deformed BF Theory}\label{section:mm}
To preface our exposition, we begin by recalling the MacDowell-Mansouri formulation of general relativity. Consider a four-dimensional Lorentzian (respectively Riemannian) manifold with boundary $(M,g)$ and let Latin letters from the middle of the alphabet,  $i, \ldots, j \in \{0,\ldots,3\} \subset \mathbb{Z}^{+}$, denote Minkowskian (Euclidean) internal indices (\emph{c.f.} \cite{wise}), corresponding to the structure group $SO(1,3)$ ($SO(4)$) of the orthonormal frame bundle.  Similarly, let Latin letters from the beginning of the alphabet, $a, \ldots, b \in \{0,\ldots,3\} \subset \mathbb{Z}^{+}$, denote the usual spacetime indices.  The key feature of the MacDowell-Mansouri formulation of general relativity is the incorporation of both the coframe $e^{i}$ and general spin connection $A^{ij}$ variables into a single Lie algebra-valued Cartan connection $\omega$ called the MacDowell-Mansouri connection \cite{wise},  
\begin{equation}\label{eq:mmconncoordfree}
\omega = A + \frac{1}{l} e.
\end{equation}
The Lie algebra $\mathfrak{g}$, in which $\omega$ takes values, is called the MacDowell-Mansouri gauge algebra and for Lorentzian or Riemannian spacetimes: $\mathfrak{g} \supset \mathfrak{so}(1,3)$ or $\mathfrak{g} \supset \mathfrak{so}(4)$, respectively.  One may equivalently write \eqref{eq:mmconncoordfree} as a $5 \times 5$ matrix
\begin{equation}\label{eq:EH-mmconnalt}
\omega^{IJ} = \left[
\begin{array}{cc}
A^{ij} & \frac{1}{l} e^{i} \\
- \frac{\epsilon}{l} e^{j} & 0
\end{array}{}
\right],
\end{equation}
where the parameter $l$ is a length scale and $\epsilon \in \{ -1,1 \} \subset \mathbb{Z}$ is determined by the choice of gauge algebra according to: \newpage
\begin{table}[!htbp]
\centering
\caption{Relation of gauge algebras to $\epsilon$}\label{table:gaugegroupstable}
\begin{tabular}{ccr}
  \toprule[1.5pt]
  Lorentzian & Riemannian & $\epsilon \p{.}$ \\
  \midrule
$\mathfrak{so}(2,3)$ & $\mathfrak{so}(1,4)$  & $-1$ \vspace{1mm} \\
$\mathfrak{so}(1,4)$ & $\mathfrak{so}(5)$ & $1$ \vspace{1mm} \\
  \bottomrule[1.5pt]\end{tabular}
\end{table}
\p{.} \\
To see why this is so, consider the fundamental representation of the Lie algebras $\mathfrak{so}(2,3)$ or $\mathfrak{so}(1,4)$ (for $\epsilon = -1$ or $1$ in the Lorentzian theory, respectively).  Wise \cite{wise} elucidates that these are $5 \times 5$ matrices of the form
\begin{equation}\label{eq:fundamentalrep}
\left[ \begin{array}{ccccc}
 0 & b^{1} & b^{2} & b^{3} & p^{0}/l  \\
b^{1} & 0 & j^{3} & -j^{2} & p^{1}/l \\
b^{2} & - j^{3}  & 0 & j^{1} & p^{2}/l \\
b^{3} & j^{2}  & - j^{1} & 0 & p^{3}/l \\
\epsilon p^{0}/l  & - \epsilon p^{1}/l & - \epsilon p^{2}/l  & - \epsilon p^{3}/l & 0  
 \end{array} \right],
\end{equation}
where the $j$'s correspond to rotations, the $b$'s to boosts and the $p$'s to translations.  By inspecting equation \eqref{eq:fundamentalrep}, one notices how appropriate choice of $\epsilon$ determines either $\mathfrak{so}(2,3)$ or $\mathfrak{so}(1,4)$. The associated curvature 2-form of \eqref{eq:EH-mmconnalt} is readily shown to be
\begin{equation}\label{eq:EH-Fmatrix}
F^{IJ} = \left[ \begin{array}{cc}
\mathsf{R}^{ij} - \frac{\epsilon}{l^{2}} e^{i} \wedge e^{j} & \frac{1}{l} \mathsf{T}^{i} \\
- \frac{\epsilon}{l} \mathsf{T}^{j} & 0
\end{array}
\right],
\end{equation}
where $\mathsf{R}^{ij}$ and $\mathsf{T}^{i}$  are the curvature and torsion 2-forms, respectively.  In equations \eqref{eq:EH-mmconnalt} and \eqref{eq:EH-Fmatrix}, the internal indices $i,\ldots,j$ correspond to the rank 4 structure groups of the frame bundle, whilst the extended set of indices $I,\ldots,J$ correspond to the MacDowell-Mansouri gauge group $G$ and instead range over $0,\ldots,4$, since $\text{rank}(G) = 5$ for any choice of $G$ the Lie group associated to the Lie algebras from table \ref{table:gaugegroupstable}. With the MacDowell-Mansouri connection and its curvature in hand, one identifies the parameter $\epsilon$ and the length scale $l$ with the cosmological constant $\Lambda$ according to
\begin{equation}\label{eq:cosmoassociation}
\frac{\Lambda}{3} := \frac{\epsilon}{l^{2}},
\end{equation}
so that the top-left block of \eqref{eq:EH-Fmatrix} becomes $F^{ij} = \mathsf{R}^{ij} - \frac{\Lambda}{3} e^{i} \wedge e^{j}$.  We refer to this block as the primary part of $F$, which we denote by the notation $\wh{F}^{IJ} = F^{ij}$. Notice in equation \eqref{eq:cosmoassociation} that the sign of the cosmological constant is in fact determined by the gauge algebra, since choosing $\mathfrak{g}$ fixes the parameter $\epsilon$.  Upon making this identification, one then constructs the MacDowell-Mansouri action as the wedge product of the top-left block of \eqref{eq:EH-Fmatrix} and its dual according to \cite{mm}
\begin{equation}\label{eq:EH-mmaction}
S_{\text{MM}} := -\frac{3}{4 G_{\text{N}} \Lambda} \int_{M} F^{ij} \wedge F^{kl} \epsilon_{ijkl},
\end{equation}
where $G_{\text{N}}$ is Newton's constant and $\epsilon_{ijkl}$ is the alternating (Levi-\v{C}ivit\`{a}) tensor.  It is interesting to note that the cosmological constant is an essential feature of MacDowell-Mansouri gravity.  To see this, notice that if $\Lambda = 0$ then the top-left block of the curvature of the MacDowell-Mansouri connection \eqref{eq:EH-Fmatrix} reduces to $F^{ij} = \mathsf{R}^{ij}$.  Therefore, the MacDowell-Mansouri action \eqref{eq:EH-mmaction} (with the prefactor of $\Lambda$ omitted) reads\footnote{In this case, the gauge algebra is the Lie algebra of the (semi-)Euclidean group: $\mathfrak{iso}(1,3)$ or $\mathfrak{iso}(4)$.}:
\begin{equation}\label{eq:eulerchar}
\chi = -\frac{3}{4 G_{\text{N}}} \int_{M} \mathsf{R}^{ij} \wedge \mathsf{R}^{kl} \epsilon_{ijkl}.
\end{equation}
This is the well-known Euler characteristic of the manifold, and is a purely topological term.  As such, the variation of \eqref{eq:eulerchar} yields no equations of motion and one does not recover general relativity.  For $\Lambda \neq 0$ however, one may readily show that \eqref{eq:EH-mmaction} is equivalent to the Palatini action
\begin{equation}
S_{\text{Pal}} = \frac{1}{2G_{\text{N}}} \int_{M} \left( e^{i} \wedge e^{j} \wedge \mathsf{R}^{kl} - \frac{\Lambda}{6} e^{i} \wedge e^{j} \wedge e^{k} \wedge e^{l} \right) \epsilon_{ijkl}
\end{equation}
on manifolds with boundary, which may be subsequently written in terms of the metric $g$ and a general connection $\Gamma$ as:
\begin{equation}\label{eq:palatiniamendment}
S_{\text{Pal}} =  \frac{1}{G_{\text{N}}} \int_{M} \ud^{4} x~  \sqrt{|g|} \Big( \mathsf{scal}(\Gamma) - 2 \Lambda \Big).
\end{equation}
By varying \eqref{eq:palatiniamendment} with respect to the connection, we find that $\Gamma$ is constrained to be the Christoffel symbol of the Levi-\v{C}ivit\`{a} connection:
\begin{equation}\label{eq:torsionfreeconstraint}
\Gamma^{c}_{\p{c}ab} = \bigg\{ \begin{array}{c} c \\ a ~~~ b  \end{array} \bigg\} := \frac{1}{2} g^{cd} \left( \partial_{a} g_{bd} + \partial_{b} g_{ad} - \partial_{d} g_{ab} \right),
\end{equation}
while variation with respect to $g$ yields the vacuum Einstein equations with a cosmological constant:
\begin{equation}\label{eq:EH-efesmm-gr}
\mathsf{Ric}_{ab}(\{ ~ \}) - \frac{1}{2} \mathsf{scal}(\{ ~ \}) g_{ab} + \Lambda g_{ab} = 0.
\end{equation}
One may equivalently write \eqref{eq:EH-efesmm-gr} as $\mathsf{Ric}_{ab} = \Lambda g_{ab}$ to emphasise that the metric $g_{ab}$ that solves the field equations is an Einstein metric.  The MacDowell-Mansouri action may be extended to higher dimensions by incorporating an additional coframe for each extra dimension \cite{vasiliev}:
\begin{equation}\label{eq:ndMM}
S_{\text{MM}}^{(n)} := -\frac{3}{4 \wt{G}_{\text{N}} \Lambda} \int_{M^{n}}  \epsilon_{i_{1} \ldots i_{n}} F^{i_{1}i_{2}} \wedge F^{i_{3}i_{4}} \wedge \underbrace{e^{i_{5}} \wedge \ldots \wedge e^{i_{n}}}_{n-4 \text{ times}},
\end{equation}
where $\wt{G}_{\text{N}}$ is Newton's constant in $n$ dimensions.  In this case, one finds upon variation that $\mathsf{Ric}_{ab} = (2/n-2) \Lambda g_{ab}$ and the gauge algebras are the obvious $n$-dimensional counterparts of the gauge algebras in table \ref{table:gaugegroupstable}, namely:
\begin{table}[!htbp]
\centering
\caption{$n$-dimensional gauge algebras} \label{table:ndgaugealgebrastable}
\vspace{1mm}
\begin{tabular}{ccr}
  \toprule[1.5pt]
  Lorentzian & Riemannian & $\epsilon \p{,}$  \\
  \midrule
$\mathfrak{so}(2,n-1)$ & $\mathfrak{so}(1,n)$  & $-1$ \vspace{1mm} \\
$\mathfrak{so}(1,n)$ & $\mathfrak{so}(n+1)$ & $1$ \vspace{1mm} \\
  \bottomrule[1.5pt]
\end{tabular}
\end{table}\\
From the curvature 2-form $F^{IJ}$ of the MacDowell-Mansouri connection, one may construct the deformed BF theory action for general relativity:
\begin{equation}\label{eq:MM-bfaction}
S_{\text{BF}} = -\int_{M} B_{IJ} \wedge F^{IJ} - \frac{G_{\text{N}} \Lambda}{12} \wh{B}_{IJ} \wedge \wh{B}_{KL} \epsilon^{IJKL},
\end{equation}
where $B_{IJ}$ is an auxiliary bivector. It's matrix form is entirely unconstrained,
\begin{equation}\label{eq:MM-b}
B_{IJ} = \left[
\begin{array}{cc}
B_{ij} & B_{i5} \\
B_{5j} & B_{55}
\end{array}{}
\right],
\end{equation}
since all but the primary part of the bivector and its dual will be seen to be absent in our results.  The theory is so-called because it is a based on a topological field-theoretic action
\begin{equation}\label{eq:tft}
S_{\text{TFT}} = - \int_{M} \text{tr} \left( B \wedge F \right),
\end{equation}
which has been ``deformed'' by adding a $B \wedge \star B$ term (\emph{c.f.} \cite{freidel}).  By itself, \eqref{eq:tft} has no propagating degrees of freedom, which motivates its deformation in the hopes of finding a tractable route towards quantum gravity.  Indeed, varying the deformed BF theory action, a short calculation shows that when the equations of motion are satisfied - \emph{i.e.} when the theory is \emph{on shell} - then the field equations so-derived are equivalent to the Einstein equations.

\section{The fundamental length scale as a dynamical variable}
With the aforementioned background in place, let us now promote the length scale $l$ in \eqref{eq:cosmoassociation} to a dynamical variable and derive the field equations from the resulting deformed BF theory action. Rather than consider $\Lambda = 10^{-52}$  $m^{-2}$ (\cite{barrow}) to be a dimensionful constant as is usual, we consider instead a dimensionless (and without loss of generality, strictly positive) $\Lambda$ whose sign is controlled by the prefactor $\epsilon$.   Therefore, choosing a particular (dimensionless) scaling of $\sqrt{\Lambda/3}$, we write the modified Cartan connection as:
\begin{equation}\label{eq:ourmodifiedconnection}
\omega^{IJ} = \left[
\begin{array}{cc}
A^{ij} &  \sqrt{\frac{\Lambda}{3}} \phi e^{i} \\
- \epsilon \sqrt{\frac{\Lambda}{3}} \phi e^{j} & 0
\end{array}{}
\right].
\end{equation}
It is clear by inspection that the prefactor of the coframe has the requisite dimension of inverse-length (\emph{c.f} \cite{wise}) subject to to this expedient choice of scaling.  Calculating the curvature of the connection \eqref{eq:ourmodifiedconnection}, we find
\begin{equation}\label{eq:confEH-Fmatrix}
F^{IJ} = \left[ \begin{array}{cc}
\mathsf{R}^{ij} - \frac{\epsilon \Lambda}{3} \phi^{2} e^{i} \wedge e^{j} & \sqrt{\frac{\Lambda}{3}} \phi \mathsf{T}^{i} + \{ \partial \phi \} \\
- \epsilon \sqrt{\frac{\Lambda}{3}} \phi \mathsf{T}^{j} + \{ \partial \phi \} & 0
\end{array}
\right],
\end{equation}
where $\mathsf{T}^{i} = \ud e^{i} + A^{i}_{\p{i}j} \wedge e^{j}$ is the torsion 2-form and $\{ \partial \phi \}$ denotes derivatives of $\phi$. With $B$ as in equation \eqref{eq:MM-bfaction}, we consider the deformed BF action
\begin{equation}\label{eq:confMM-reducedbf}
S^{\phi}_{\text{BF}} = -\int_{M} \phi^{2} B_{IJ} \wedge F^{IJ} - \frac{\epsilon G_{\text{N}} \Lambda}{12} \phi^{4} \wh{B}_{IJ} \wedge \wh{B}_{KL} \epsilon^{IJKL}.
\end{equation}
Let us vary the action \eqref{eq:confMM-reducedbf} in a piecewise fashion with respect to the physical fields $\omega$, $B$ and $\phi$, such that the total variation is given by the sum $\delta S^{\phi}_{\text{BF}} = \delta_{B} S^{\phi}_{\text{BF}} + \delta_{\omega} S^{\phi}_{\text{BF}} + \delta_{\phi} S^{\phi}_{\text{BF}}$.  To calculate the variation with respect to the $B$ field,
\begin{equation}\label{eq:confMM-variationb}
\delta_{B} S^{\phi}_{\text{BF}} = -\int_{M} \phi^{2} \delta B_{IJ} \wedge F^{IJ} - \frac{ \epsilon G_{\text{N}} \Lambda}{12} \phi^{4} \delta \left( B_{IJ} \wedge \wh{B}_{KL} \epsilon^{IJKL} \right),
\end{equation}
we use symmetry of the Hodge star on 2-forms, the total antisymmetry of the alternating tensor, and substitute the resulting expression into \eqref{eq:confMM-variationb} to find that
\begin{equation}
\delta_{B} S^{\phi}_{\text{BF}} = -\int_{M} \phi^{2} \delta B_{IJ} \wedge \left( F^{IJ} - \frac{\epsilon G_{\text{N}} \Lambda}{6} \phi^{2} \wh{B}_{KL} \epsilon^{IJKL} \right).
\end{equation}
Next, for the variation with respect to $\omega$,
\begin{equation}\label{eq:confMM-varyf}
\delta_{\omega} S^{\phi}_{\text{BF}} = -\int_{M} \phi^{2} B_{IJ} \wedge \delta F^{IJ},
\end{equation}
we recall that the expression exterior covariant derivative of a $2$-form $\sigma$ is given by $\ud_{\omega} \sigma^{IJ} := \ud \sigma^{IJ} + \omega^{I}_{\p{I}K} \wedge \sigma^{KJ}$, and so for $F^{IJ} = \ud_{\omega} \omega^{IJ}$ we have that $\delta F^{IJ} = \delta \left( \ud_{\omega} \omega^{IJ} \right) = \ud_{\omega} \delta \omega^{IJ}$.  Substituting this expression into \eqref{eq:confMM-varyf} and integrating by parts, we find
\begin{equation*}
\delta_{\omega} S^{\phi}_{\text{BF}} = -\int_{M} \phi^{2} \delta \omega^{IJ} \wedge \ud_{\omega} B_{IJ}.
\end{equation*}
Finally, for the variation with respect to $\phi$,
\begin{equation}\label{eq:confMM-varyphi}
\delta_{\phi} S^{\phi}_{\text{BF}} = -\int_{M} \delta \phi^{2} B_{IJ} \wedge F^{IJ} - \frac{\epsilon G_{\text{N}} \Lambda}{12} \delta \phi^{4} \wh{B}_{IJ} \wedge \wh{B}_{KL} \epsilon^{IJKL},
\nonumber
\end{equation}
we use the fact that $\delta (f (\phi)) = \partial / \partial \phi \left( f(\phi) \right) \cdot \delta \phi$ for some function $f(\phi)$, and so \eqref{eq:confMM-varyphi} reduces to
\begin{equation}
\delta_{\phi} S^{\phi}_{\text{BF}} = -\int_{M} \delta \phi \left( 2 \phi B_{IJ} \wedge F^{IJ} - \frac{\epsilon G_{\text{N}} \Lambda}{3} \phi^{3} \wh{B}_{IJ} \wedge \wh{B}_{KL} \epsilon^{IJKL} \right).
\nonumber
\end{equation}
Consequently, the equations of motion are
\begin{align}
F^{IJ} &= \frac{\epsilon G_{\text{N}} \Lambda}{6} \phi^{2} \wh{B}_{KL} \epsilon^{IJKL} \label{eq:confMM-eoms1}, \\
\ud_{\omega} B_{IJ} &= 0 \label{eq:confMM-eoms2},\\
2 \phi B_{IJ} \wedge F^{IJ} &= \frac{\epsilon G_{\text{N}} \Lambda}{3} \phi^{3} \wh{B}_{IJ} \wedge \wh{B}_{KL} \epsilon^{IJKL} \label{eq:confMM-eoms3a}.
\end{align}
Notice in equation \eqref{eq:confMM-eoms1} that the full $F$ tensor is proportional to the primary part $\wh{B}$ of $B$, so it follows that  $\wh{F} \propto \star \wh{B}$ for the primary part of $F$. Likewise, the primary parts $\wh{F}$ and $\wh{B}$ contain solely four-dimensional (lower case Latin) indices, and so \eqref{eq:confMM-eoms1} and \eqref{eq:confMM-eoms3a} reduce to
\begin{align}
F^{ij} &= \frac{\epsilon G_{\text{N}} \Lambda}{6} \phi^{2} B_{kl} \epsilon^{ijkl} \label{eq:confMM-eoms},\\
2 \phi B_{ij} \wedge B^{ij} &= \frac{\epsilon G_{\text{N}} \Lambda}{3} \phi^{3} B_{ij} \wedge B_{kl} \epsilon^{ijkl} \label{eq:confMM-eoms3},
\end{align}
where $B = \wh{B}$ is imposed by consistency of the index structure.  Consequently, we may write the deformed BF theory action as 
\begin{equation}\label{eq:confMM-cbf}
S^{\phi}_{\text{BF}} = -\int_{M} \phi^{2} B_{ij} \wedge F^{ij} - \frac{\epsilon G_{\text{N}} \Lambda}{12} \phi^{4} B_{ij} \wedge B_{kl} \epsilon^{ijkl}.
\end{equation}
Let us now show that the third equation of motion, \eqref{eq:confMM-eoms3}, is always satisfied by virtue of the first, \eqref{eq:confMM-eoms}.  We solve \eqref{eq:confMM-eoms1} for $B$ by using the fact that the square of the Hodge star operator on a $2$-form $\sigma$ on a Lorentzian manifold acts according to $\star \star \sigma = - \sigma$, and thus find that $B_{ij} = - (3 / 2\epsilon G_{\text{N}} \Lambda \phi^{2}) \epsilon_{ijkl} F^{kl}$.  Substituting this into \eqref{eq:confMM-eoms3}, we find
\begin{equation*}
0 = 2 \phi B_{ij} \wedge F^{ij} - \frac{\epsilon G_{\text{N}} \Lambda}{3} \phi^{3} B_{ij} \wedge B_{kl} \epsilon^{ijkl} = \frac{3}{\epsilon G_{\text{N}} \Lambda} F^{ij} \wedge F^{kl} \epsilon_{ijkl} - \frac{3}{\epsilon G_{\text{N}} \Lambda} F^{ij} \wedge F^{kl} \epsilon_{abcd} = 0
\end{equation*}
as claimed.  Let us now show that the deformed BF action is equivalent to a MacDowell Mansouri-type action, and thence to a Palatini-type action with a quartic potential.  It follows by substitution and the anti-symmetry of the forms that 
\begin{align}\label{eq:confEH-MMrecovered}
S^{\phi}_{\text{BF}} &= -\int_{M} \phi^{2} B_{ij} \wedge F^{ij} - \frac{\epsilon G_{\text{N}} \Lambda}{12} \phi^{4} B_{ij} \wedge B_{kl} \epsilon^{ijkl} \nonumber \\
&= -\int_{M} \phi^{2} \left( - \frac{3}{2\epsilon G_{\text{N}} \Lambda \phi^{2}} \epsilon_{ijkl} F^{kl} \wedge F^{ij} \right) - \frac{\epsilon G_{\text{N}} \Lambda}{12} \phi^{4}  \left( - \frac{3}{2\epsilon G_{\text{N}} \Lambda \phi^{2}} \epsilon_{ijkl} F^{kl} \wedge \frac{6}{\epsilon G_{\text{N}} \Lambda \phi^{2}} F^{ij} \right)\nonumber \\
&= -\frac{3 \epsilon}{4G_{\text{N}} \Lambda} \int_{M} F^{ij} \wedge F^{kl} \epsilon_{ijkl} =: S^{\phi}_{\text{MM}},
\end{align}
where we have also used the fact that $\epsilon = 1 / \epsilon$.  Let us reduce this further by substituting the $\wh{F}$ component of \eqref{eq:confEH-Fmatrix} into \eqref{eq:confEH-MMrecovered}:
\begin{align}\label{eq:confEH-palworking1}
-\frac{3 \epsilon}{4G_{\text{N}} \Lambda} \int_{M} F^{ij} \wedge F^{kl} \epsilon_{ijkl} &= -\frac{3\epsilon}{4 G_{\text{N}} \Lambda} \int_{M} \mathsf{R}^{ij} \wedge \mathsf{R}^{kl} \epsilon_{ijkl} - \frac{\epsilon \Lambda}{3} \phi^{2} \mathsf{R}^{ij} \wedge e^{k} \wedge e^{l} \epsilon_{ijkl} \nonumber \\
&\p{aaaaaaaa}~~~~~~~~~~~~~~~~- \frac{\epsilon \Lambda}{3} \phi^{2} e^{i} \wedge e^{j} \wedge \mathsf{R}^{kl} \epsilon_{ijkl} + \frac{\Lambda^{2}}{9} \phi^{4} e^{i} \wedge e^{j} \wedge e^{k} \wedge e^{l} \epsilon_{ijkl}.
\end{align}
The first term is purely topological, so we disregard it.  Using the fact that the Hodge star acts on $2$-forms according to $\star \sigma_{ij} = (1/2) \epsilon_{ijkl} \sigma^{kl}$, and is summetric: $\sigma \wedge \star \tau = \tau \wedge \star \sigma$, we see that \eqref{eq:confEH-palworking1} reduces to
\begin{equation}\label{eq:confEH-twoterms}
-\frac{3 \epsilon}{4G_{\text{N}} \Lambda} \int_{M} F^{ij} \wedge F^{kl} \epsilon_{ijkl} =  \frac{1}{G_{\text{N}}} \int_{M} \frac{1}{2} \phi^{2} e^{i} \wedge e^{j} \wedge \mathsf{R}^{kl} \epsilon_{ijkl} - \frac{\epsilon \Lambda}{12} \phi^{4} e^{i} \wedge e^{j} \wedge e^{k} \wedge e^{l} \epsilon_{abcd}.
\end{equation}
Let us analyse the first of these two terms first. it follows trivially that $\frac{1}{2} \phi^{2} e^{i} \wedge e^{j} \wedge \mathsf{R}^{kl} \epsilon_{ijkl} =  \phi^{2} e^{i} \wedge e^{j} \wedge \star \mathsf{R}_{ij} = \phi^{2} \mathsf{R}_{ij} \wedge \star (e^{i} \wedge e^{j})$.  Now, changing the dummy indices for ease of calculation, the Riemann tensor 2-form expands to $\phi^{2} \mathsf{R}_{jk} \wedge \star (e^{j} \wedge e^{k}) = \frac{1}{2} \phi^{2} \mathsf{R}_{jkil} e^{i} \wedge e^{l} \wedge \star (e^{j} \wedge e^{k}).$.  Recalling that $\tau \wedge \star \sigma = \langle \tau, \sigma \rangle \star 1$, then we have
\begin{equation}\label{eq:confEH-scalterm}
\frac{1}{2} \phi^{2} \mathsf{R}_{jkil} e^{i} \wedge e^{l} \wedge \star (e^{j} \wedge e^{k}) = \frac{1}{2} \phi^{2} \mathsf{R}_{jkil} \left( \eta^{ik} \eta^{jl} - \eta^{il} \eta^{jk} \right) \star 1 = \phi^{2} \mathsf{scal}(A) \sqrt{-g}~ \ud^{4}x,
\end{equation}
where the middle equality follows by some lengthy tensor algebra, and $A = \wh{\omega}$ is the primary part of the MacDowell-Mansouri connection. Now, recalling that $\star 1 = \frac{1}{4!} \epsilon_{ijkl} e^{i} \wedge e^{j} \wedge e^{k} \wedge e^{l}$ then the second term in \eqref{eq:confEH-twoterms} becomes
\begin{equation}\label{eq:confEH-phifourterm}
- \frac{\epsilon\Lambda}{12} \phi^{4} e^{i} \wedge e^{j} \wedge e^{k} \wedge e^{l} \epsilon_{ijkl} = - \frac{4! \cdot \epsilon \Lambda}{12}\phi^{4} \star 1 = - 2\epsilon \Lambda \phi^{4} \sqrt{-g}~ \ud^{4}x.
\end{equation}
Consequently, the terms \eqref{eq:confEH-scalterm} and \eqref{eq:confEH-phifourterm} combine to give
\begin{equation}\label{eq:confEH-reducedaction}
S^{\phi}_{\text{MM}} = \frac{1}{G_{\text{N}}} \int_{M} \ud^{4}x~ \sqrt{-g} \left( \phi^{2} \mathsf{scal}(A) - 2\epsilon \Lambda \phi^{4}  \right).\\
\end{equation}
In their article \cite{burtonmann}, Burton and Mann outline the procedure for varying Palatini-type actions for scalar-tensor theories, where one treats the connection $A$ and the metric $g$ as independent variables.  Let us summarise one of their main results (\emph{c.f.} \cite{burtonmann} for full details).  Consider the general action functional
\begin{equation}\label{eq:burtonmanngeneralaction}
S = \int_{M^{n}} \ud^{n}x~ \sqrt{-g}~ \bigg( D(\phi) \mathsf{scal}(A) + C(\phi) \left( \nabla^{A} \phi \right)^{2}  \bigg)
\end{equation}
for a scalar-tensor theory on an $n$-dimensional Lorentzian manifold $M^{n}$, where $\nabla^{A}$ is the covariant derivative associated to the connection $A$.  Burton and Mann give that the variation of \eqref{eq:burtonmanngeneralaction} with respect to the metric $g$ and scalar field $\phi$ results in the (vacuum) equations of motion:
\begin{align}
0 &= D(\phi) \mathsf{G}_{ab}\left( \{~ \} \right)  - \Big( D''(\phi) - C(\phi) + Q(\phi) \Big) \partial_{a} \phi \partial_{b} \phi \nonumber \\
&~~~~~ + \left( D''(\phi) - \frac{1}{2} C(\phi) + \frac{1}{2} Q(\phi)  \right) g_{ab} \partial_{c} \phi \partial^{c} \phi - D'(\phi) \Big( \nabla_{a} (\partial_{b} \phi) - g_{ab} \square \phi \Big)  \label{eq:burtonmannceheom1}
\end{align}
and
\begin{equation}\label{eq:burtonmannceheom2}
0 = D'(\phi)  \mathsf{scal}\left( \{~ \} \right) + \left( Q'(\phi) - C'(\phi) \right) g^{ab} \partial_{a} \phi \partial_{b} \phi + 2 \left( Q(\phi) - C(\phi) \right) \square \phi
\end{equation}
respectively, where
\begin{equation}\label{eq:qandx}
Q(\phi) = \frac{(n-1)(2-n)}{4}  D(\phi) X^{2}(\phi), ~~~ \text{and} ~~~ X(\phi) = \frac{2}{n-2} \frac{D'(\phi)}{D(\phi)},
\end{equation}
and $\nabla$ is the covariant derivative associated to the Levi-\v{C}ivit\`{a} connection. For the variation with respect to the connection, they give that the equation of motion is:
\begin{equation}\label{eq:burtonmannconnscalar}
\Gamma^{a}_{\p{a}bc} = \left\{ \begin{array}{c} a \\ b ~~~ c  \end{array} \right\} + \frac{1}{2} X(\phi) \Big( (\partial_{b} \phi) \delta^{a}_{c} + (\partial_{c} \phi) \delta^{a}_{b} - (\partial^{a} \phi) g_{bc} \Big).
\end{equation}
They term such a variation - where the connection is not \emph{a priori} fixed - a Palatini variation, in contrast to a Hilbert variation where the connection is mandated to be the Levi-\v{C}ivit\`{a} connection.  A short calculation shows that equation \eqref{eq:burtonmannconnscalar} is the image of the conformal transformation of the Christoffel symbol for $g_{ab} \rightarrow e^{2 \Upsilon} g_{ab},$ with the conformal factor identified with the scalar field, $e^{\Upsilon} = \phi$.   Our $n=4$ action \eqref{eq:confEH-reducedaction} corresponds to $D(\phi) = \phi^{2}$ and $C(\phi) = 0$ in \eqref{eq:burtonmanngeneralaction}, and so substituting these values into \eqref{eq:qandx}, we have that $Q = -6$ and $X(\phi) = 2 / \phi$.  Therefore, using equations \eqref{eq:burtonmannceheom1} and \eqref{eq:burtonmannceheom2}, the (vacuum) equations of motion obtained by varying \eqref{eq:confEH-reducedaction} are
\begin{equation}
\bigg\{ \begin{array}{l}
0 = \phi^{2} \mathsf{G}_{ab}(\{~ \}) + 4 \partial_{a} \phi \partial_{b} \phi - g_{ab} \partial_{c} \phi \partial^{c} \phi - 2 \phi \left( \partial_{a} \partial_{b}  - g_{ab} \square \right) \phi + g_{ab} \Lambda \phi^{4},\\
0 = 2 \phi \mathsf{scal}(\{~ \}) - 12 \square \phi - 8 \Lambda \phi^{3},
\end{array}
\end{equation}
These are precisely the (vacuum) conformal Einstein equations \eqref{eq:cefestogether} obtained by varying the conformal Einstein-Hilbert action:
\begin{equation}
S_{\text{CEH}} :=  \int_{M} \ud^{4} x~ \sqrt{-g}  \Big( \phi^{2} \mathsf{scal}(\{ ~ \})  + 6 g^{ab} \partial_{a} \phi \partial_{b} \phi - 2 \Lambda \phi^{4} \Big),
\end{equation}
thus proving our assertion that considering the fundamental length scale of the MacDowell-Mansouri as a dynamical variable transforms \eqref{eq:EH-mmaction} into an action for conformal gravity.
.

\section{The geometrical origin of the scalar field}
Casting conformal gravity in the form of a gauge theory endows the scalar field with a geometrical description.  To show this however, we must introduce some technology from Cartan geometry.  Our principal source for this review is Derek Wise's excellent article \emph{MacDowell-Mansouri gravity and Cartan geometry} \cite{wise}, which thoroughly elucidates the subject.  To set some notation, consider (for example) the $p$\textsuperscript{th} exterior power of the cotangent bundle, $\Lambda^{p} T^{*}M$.  Recall that the space of $p$-forms on $M$ is the space of sections of the fibre bundle $\Lambda^{p} T^{*}M \to M$,
\begin{equation}
\Omega^{p} \left( M \right) := \Gamma \left( \Lambda^{p} T^{*}M \right).
\end{equation}
Consequently, a $p$-form $\eta$ with values in (say) a Lie algebra $\mathfrak{g}$ is an element of the space of sections of the product bundle $\eta \in \Gamma \left( \Lambda^{p} T^{*}M \otimes \mathfrak{g} \right)$, which we write in a more condensed fashion as $\eta \in \Omega^{p} \left( M, \mathfrak{g} \right)$.  Now, let $\mathcal{P}^{g}$ denote the bundle of $g$-orthonormal frames on $M$.  We say that an Ehresmann connection form on the frame bundle is a 1-form
\begin{equation}\label{eq:confhol-ehresmann}
A \in \Omega^{1} \left(\mathcal{P}^{g}, {\mathfrak{h}} \right), 
\end{equation}
where $\mathfrak{h}$ is the Lie algebra of the structure group $H$.  For gravity, $H = SO(1,3)$ or $SO(4)$ in Lorentzian or Riemannian signature, respectively.  An Ehresmann connection form satisfies the properties
\begin{enumerate}
\item{$R^{*}_{h}A = \text{Ad}(h^{-1})A$ for all $h \in H$ and}
\item{$A$ restricts to the Maurer-Cartan form $A_{H}$ on the fibres of $\mathcal{P}^{g}$,}
\end{enumerate}
where $R^{*}$ is the pullback of the right action of $H$ on $\mathcal{P}^{g}$ and $A_{H} : T\mathcal{P}^{g}_{x} \to \mathfrak{h}$ is the Maurer-Cartan form on the fibres of $\mathcal{P}^{g}$.  Less familiar in theoretical physics is the notion of a Cartan connection form (of type $G$) on the frame bundle\footnote{It is important to note that our symbols for the Ehresmann and Cartan connections are the opposite to those in references \cite{wise} and \cite{freidel}.}:
\begin{equation}
\omega \in \Omega^{1} \left(\mathcal{P}^{g}, {\mathfrak{g}} \right), 
\end{equation}
which takes values in the ambient gauge algebra $\mathfrak{g} \supset \mathfrak{h}$.  The Cartan connection form satisfies the conditions
\begin{enumerate}
\item{$R^{*}_{h} \omega = \text{Ad}(h^{-1}) \omega$ for all $h \in H$,}
\item{$\omega(\wt{X}) = X$ for all $X \in \mathfrak{h}$ and}
\item{$\omega_{p} : T_{p}\mathcal{P}^{g} \to \mathfrak{g}$ is an isomorphism for all $p \in \mathcal{P}^{g}$,}
\end{enumerate}
where $\wt{X}$ is a fundamental vector field on $\mathcal{P}^{g}$. The gauge algebra $\mathfrak{g}$ admits, as a vector space. the decomposition
\begin{equation}\label{eq:confhol-killingdecomp}
\mathfrak{g} = \mathfrak{h} \oplus \mathfrak{g}/\mathfrak{h},
\end{equation}
which is orthogonal with respect to the Killing form on the Lie algebra.  When the decomposition \eqref{eq:confhol-killingdecomp} is such that each summand is an $\text{Ad}(H)$ representation, then \eqref{eq:confhol-killingdecomp} is called a reductive decomposition (of $\mathfrak{g}$).  Lie algebras which admit reductive decompositions are naturally called reductive Lie algebras, and indeed, all of the $n$-dimensional MacDowell-Mansouri gauge algebras in table \ref{table:ndgaugealgebrastable} are reductive.  Wise comments that only reductive Cartan connections $\omega$ (those that take values in reductive Lie algebras) may be decomposed into spin connection and coframe parts.  To effect this decomposition, consider maps $\Pi_{1}$ and $\Pi_{2}$ that project the ambient Lie algebra to each of its summands,  
\begin{equation}\label{eq:sosplitting}
\xymatrix{
& \ar[dl]_{\Pi_{1}} \ar[dr]^{\Pi_{2}} \mathfrak{g} &\\
\mathfrak{h} & & \mathfrak{g}/\mathfrak{h}.}
\end{equation}
By regarding $\omega$ itself as a map, $\omega : T\mathcal{P}^{g} \longrightarrow \mathfrak{g}$, the spin connection and coframe fields are compositely constructed from these projectors:
\begin{align}\label{eq:confhol-pullbacks}
\Pi_{1} \circ \omega &= A: T\mathcal{P}^{g} \longrightarrow \mathfrak{h}, \nonumber \\
\Pi_{2} \circ \omega &= \frac{1}{l}e : T\mathcal{P}^{g} \longrightarrow \mathfrak{g}/\mathfrak{h},
\end{align}
so that $\omega = A + (1/l) e$ by virtue of equation \eqref{eq:confhol-killingdecomp}.  This reductive decomposition \eqref{eq:confhol-killingdecomp} is what allows the spin connection and coframe fields to be regarded as Lie algebra-valued forms $A \in \Omega^{1}(\mathcal{P}^{g}, \mathfrak{h})$ and $e \in \Omega^{1}(\mathcal{P}^{g}, \mathfrak{g}/\mathfrak{h})$ on the frame bundle $\mathcal{P}^{g}$, rather than as 1-forms on spacetime \cite{durka}:
\begin{equation}
A^{ij} = A^{ij}_{\p{ij}a} \ud x^{a}, ~~~~~~ e^{i} = e^{i}_{\p{i}a} \ud x^{a}.
\end{equation}
Consequently, we may regard the curvature of the MacDowell-Mansouri connection as a Lie algebra-valued form $F \in \Omega^{2}(\mathcal{P}^{g}, \mathfrak{g})$ with primary part $\wh{F} \in \Omega^{2}(\mathcal{P}^{g}, \mathfrak{h})$. To see that the gauge-theoretic action
\begin{equation}\label{eq:mmactioncoordfree}
S_{\text{MM}} = -\frac{3}{2 G_{\text{N}} \Lambda} \int_{M} \text{tr} \left( \wh{F} \wedge \star \wh{F} \right),
\end{equation}
which is the trace of the product of the curvature 2-form of an Ehresmann connection and its dual, describes general relativity, recall that the internal Hodge star operator $\star$ on $\mathfrak{h}$ is related to the alternating (Levi-\v{C}ivit\`{a}) tensor $\epsilon_{ijkl}$ on the tangent space by
\begin{equation}\label{eq:internalhodge}
\star = - \frac{1}{2!} \epsilon_{ijkl}.
\end{equation}
The MacDowell-Mansouri action in index notation, \eqref{eq:EH-mmaction}, is then recovered by substituting $\wh{F} = F^{ij}$ and \eqref{eq:internalhodge} into \eqref{eq:mmactioncoordfree}.\\

Returning to the axioms satisfied by the Cartan connection, notice that a necessary condition for the existence of a Cartan connection (of type $G$) on an $H$-principal bundle $P \to M$ with values in $\mathfrak{g}$ is that the tangent space $T_{p}P$ at each $p \in P$ is isomorphic to $\mathfrak{g}$ itself.  This is equivalent to the requirement that $T_{x} M \cong \mathfrak{g}/\mathfrak{h}$ at each $x \in M$, so spacetime must have the same dimension as the homogeneous space $G/H$, which is called a tangent model geometry in Cartan geometry.  The tangent model geometry $G/H$ is endowed with a canonical metric of constant scalar curvature, the sign of the scalar curvature of which matches that of the Einstein metric $g$ on the Einstein space $(M,g)$ obtained by the procedure outlined in section \ref{section:mm}.  Since the signs of the scalar curvatures of the spacetime and homogeneous space metrics match, infinitesimally approximating $M$ by $G/H$ is regarded as more natural than infinitesimal approximation by a tangent space (the latter approximation, Wise comments, is implicit in all other formulations of general relativity \cite{wise}).  The radius of the homogeneous space $G/H$ is the length scale $l$ in the MacDowell-Mansouri connection, so we see that a notable geometrical difference between general relativity and its conformally transformed counterpart is that the radius of the homogeneous space to which spacetime is infinitesimally approximated varies as a scalar field in the latter, but remains fixed in the former.  The positivity of the radius is ensured by the positivity of the conformal factor.

\end{document}